\documentclass[pra,a4paper,superscriptaddress,twocolumn,showpacs,amsmath,amssymb,floatfix]{revtex4}

\usepackage{graphicx}
\usepackage{amssymb}
\usepackage{float}
\usepackage{hyperref}
\hypersetup{
	bookmarks=true,         % show bookmarks bar?
	unicode=false,          % non-Latin characters in Acrobat’s bookmarks
	pdftoolbar=true,        % show Acrobat’s toolbar?
	pdfmenubar=true,        % show Acrobat’s menu?
	pdffitwindow=false,     % window fit to page when opened
	pdfstartview={FitH},    % fits the width of the page to the window
	pdftitle={My title},    % title
	pdfauthor={Author},     % author
	pdfsubject={Subject},   % subject of the document
	pdfcreator={Creator},   % creator of the document
	pdfproducer={Producer}, % producer of the document
	pdfkeywords={keyword1, key2, key3}, % list of keywords
	pdfnewwindow=true,      % links in new PDF window
	colorlinks=true,       % false: boxed links; true: colored links
	linkcolor=blue,          % color of internal links (change box color with linkbordercolor)
	citecolor=blue,        % color of links to bibliography
	filecolor=blue,         % color of file links
	urlcolor=blue        % color of external links
}

\usepackage{color}

\usepackage[normalem]{ ulem }
\usepackage{soul}

\input{epsf}

\begin{document}

\title{Random lasing from
Anderson attractors}

\author{Guillaume Rollin}
\affiliation{Institut UTINAM, UMR 6213, CNRS, 
Université Bourgogne Franche-Comt\'e Besançon, France}
\author{Jos\'e Lages}
\affiliation{Institut UTINAM, UMR 6213, CNRS, 
Université Bourgogne Franche-Comt\'e Besançon, France}
\author{Dima L. Shepelyansky}
%\homepage[]{http://www.quantware.ups-tlse.fr}
\affiliation{\mbox{Laboratoire de Physique Th\'eorique, 
Universit\'e de Toulouse, CNRS, UPS, 31062 Toulouse, France}}

%\date{today}
\date{January 11, 2022}

\begin{abstract}
  We introduce and study
  a two-dimensional dissipative nonlinear Anderson pumping  model
  which is characterized by localized or delocalized eigenmodes in a linear
  regime and in addition includes nonlinearity, dissipation and pumping.
  We find that above a certain pumping threshold the model
  has narrow spectral lasing lines generated by isolated
  clusters of Anderson attractors. With the increase of the pumping,
  the lasing spectrum is broaden even if narrow
  lasing peaks are still well present in the localized phase of linear modes.
  In the metallic phase, the presence of narrow spectral peaks
  is significantly suppressed. We argue that the model
  captures main features observed for random lasers.
\end{abstract}

%\pacs{05.45.Mt, 67.85.Hj,  47.27.-i, 72.15.Rn}
%05.45.-a Nonlinear dynamics and chaos
%67.85.Hj 	Bose-Einstein condensates in optical potentials
%47.27.-i 	Turbulent flows
%72.15.Rn 	Localization effects (Anderson or weak localization) 
%
%47.35.-i 	Hydrodynamic waves
%47.35.Bb 	Gravity waves 
%89.75.-k 	Complex systems 

\maketitle

\section{Introduction} 
\label{sec1}

The theory of random lasing in a disordered active media
was introduced by V.S. Letokhov in 1967-1968 \cite{letokhov}.
At present, various types of random lasers
operating in different gain and scattering media,
including powder and fibers, have been experimentally realized as reviewed in \cite{cao1,cao2,lodahl,turitsyn}.
The active interest to random lasers
is stimulated not only by their technological applications
but also by a variety of inter-disciplinary links to other research fields, such as the theory
of disordered and mesoscopic systems \cite{akkermans}, Anderson localization and
transport \cite{anderson,mirlin}, nonlinear waves in disordered media \cite{kivshar,skipetrov}, 
chaotic dynamics and strange attractors \cite{lichtenberg,ott}, 
synchronization \cite{sync}, material science, spectroscopy
and laser physics (see e.g. \cite{laser}).

Due to such an interdisciplinary nature and complexity
of random lasing systems, deep theoretical studies are required with
applications of advanced analytical and numerical tools and methods.
Various numerical studies have been reported 
with the main objective to explain specific features
of random lasing observed in experiments (see e.g.
\cite{tureci,jacquod}). However, an interplay on nonlinearity,
disorder, dissipation and pumping results in a rather complex
dynamics which the properties are rather difficult to capture and 
investigate in studies of specific modeling of an experimental setup.
Due to these reasons, we introduce here a simplified 
Dissipative Nonlinear Anderson Pumping (DINAP) model 
which in various limiting regimes describes such generic
phenomena as Anderson localization, transport in disordered media,
nonlinear waves, dissipation, pumping, synchronization,
and chaotic dissipative dynamics. We show that a lasing in such a model
captures the main features of random lasers.

\section{Model description} 
\label{sec2}

The DINAP model is described by the time evolution equations
\begin{eqnarray}
i\dot{A}_{x,y}&=&\ E_{x,y} A_{x,y}+\beta\left|A_{x,y}\right|^2 A_{x,y} + \left(1-i\eta\right)\left(-A_{x,y+1}\right.\nonumber\\
 &+&\left.4A_{x,y}-A_{x,y-1}-A_{x+1,y}-A_{x-1,y}\right)\nonumber\\
 &+&i\left(\alpha-\sigma\left|A_{x,y}\right|^2\right) A_{x,y}.
 \label{eq1}
\end{eqnarray}
Here, $A_{x,y}$ is the radiation field amplitude on
the site $\left(x,y\right)$ of a $N\times N$ square lattice with periodic boundary conditions,
$E_{x,y}$ are on site unperturbed energies
randomly distributed in the $\left[-W/2,W/2\right]$ interval.
For $\beta=\eta=\alpha=\sigma=0$, the model
is reduced to the two-dimensional Anderson model (see e.g. \cite{mirlin})
with a unit hopping amplitude on nearby sites.
In absence of disorder, i.e. at $W=0$, the spectrum of linear waves 
on a lattice of size $N \times N$ has the form
$\lambda_{q_x,q_y}=4-2\cos\left(2\pi 
q_x/N\right)-2\cos\left(2\pi q_y/N\right)$,
where $q_x$ and $q_y$ are wave numbers of ballistic waves.
In presence of disorder, i.e. $W>0$, all the eigenstates are exponentially
localized but the localization length increases exponentially 
with a decrease of the disorder strength $W$ \cite{akkermans,mirlin}.
For a lattice of finite size $N \sim 100$, the eigenstates are well localized
at $W=6 - 8$ \cite{lagesbls}.

For finite $\beta$ and $\eta=\alpha=\sigma=0$, the DINAP model is reduced to the 
2D Discrete Anderson Nonlinear Schr\"odinger equation (DANSE) model 
which was actively studied for the investigation of 
effects of weak nonlinearity on the Anderson localization 
(see e.g. \cite{danse1,danse2,flach}). It was shown that a moderate
nonlinearity leads to a destruction of the localization and a subdiffusive
spreading of the field over the lattice.

The DINAP model has several new features compared to the unitary DANSE model. 
Indeed, the parameter $\alpha$ describes lasing instability of the nonlinear media
which is balanced by the linear damping $\eta$-term and the more significant nonlinear
damping $\sigma$-term. Due to that, the DINAP model captures various nontrivial features of 
the nonlinear lasing in dissipative media with disorder.
Due to nonlinearity and disorder, it is natural to expect that 
the dynamics will be characterized by the presence of chaotic attractors
which are typical for nonlinear dissipative systems \cite{lichtenberg,ott}.

We note that a similar model in 1D was studied in \cite{lapteva}.
A number of interesting results have been reported there.
In our studies, we analyze a more realistic 2D case and concentrate the
investigations on the lasing spectrum produced by  the nonlinear media
of DINAP model.

The numerical integration of the coupled equations (\ref{eq1}) is done in the frame of 
the Trotter decomposition used in \cite{danse1,danse2}. This integration scheme
is symplectic (at $\eta=\alpha=\sigma=0$) and allows to perform 
accurate numerical simulations on large time scales.
The physical arguments which explain the accuracy and the advantages of such
integration are described in \cite{ymcolors}.

In the numerical simulations, we usually use the integration time 
step, $\Delta =0.1$, checking that the variation of this step by several times
is not affecting the obtained results. The main part of the results is presented 
for the lattice size $N \times N = 128 \times 128$. Such a size is significantly
larger than the localization length of linear eigenstates with a typical disorder strength $W=8$.

At the initial time $t=0$, a field $A_{x,y}$ is taken as random with 
typical amplitudes $|A_{x,y}|^2 \approx 7 \times 10^{-11}$ with a standard deviation
being approximately $4 \times 10^{-11}$. For a fixed random configuration
of the energies $E_{x,y}$, the initial field amplitudes do not influence the field amplitudes at large times $t\sim 10^5$ (steady-state)
since the field time evolution converges to fixed 
Anderson attractors distributed on the lattice.
This Anderson attractor steady state, averaged over a moderate time interval $\Delta t \sim 10^3$,
is independent of the above described initial random field realization $A_{x,y}$.
 
The obtained numerical results are described in the next section.

\begin{figure}[t]
\begin{center}
\includegraphics[width=0.42\textwidth]{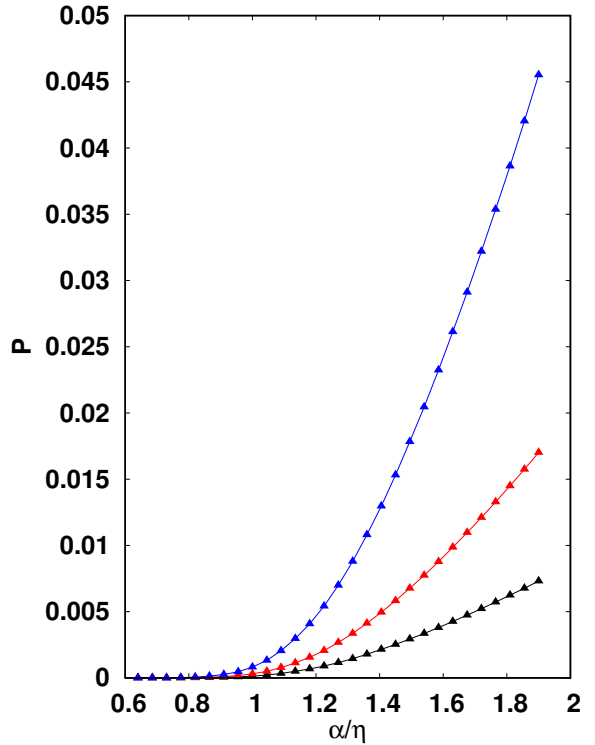}
\end{center}
\vglue -0.3cm
\caption{\label{fig1}Dependence of the space averaged steady-state field power
$P= \left<\left|A_{x,y}\right|^2 \right>$ on the active media parameter
$\alpha/\eta$.
Here, the parameters are $W=8$, $\beta=\sigma=1$. For $\eta=0.1$,
the steady-state is obtained at $t_e=10^4$
with an averaging over time interval $\Delta t =10^3$. The
lattice size is $128 \times 128$.
The values of the linear damping are $\eta=0.2$ (blue points), $\eta=0.1$ (red points), and $\eta=0.05$ (black points).
}
\end{figure}

\section{Results} 
\label{sec3}

In Fig.~\ref{fig1}, we show the dependence of the space averaged steady-state field power
$P= \left<\left|A_{x,y}\right|^2 \right>$ on the rescaled active media parameter $\alpha/\eta$.
The field growth is  generated by an effect of active media
described by the parameter $\alpha$ growth. The dissipative effects 
are produced by the $\eta$-term. The field growth is limited by the 
nonlinear dissipative $\sigma$-term.
Thus, at small values of the ratio $\alpha/\eta$, the generated field
remains small so that $P \ll 1$ for $\alpha/\eta < 0.7$.
In contrast, above the threshold value $\alpha/\eta \approx0.7$,
the field power is growing significantly
that corresponds to the random lasing regime.
We note that a similar behavior has been described  
for the 1D model \cite{lapteva}.

\begin{figure}[t]
\begin{center}
\includegraphics[width=0.42\textwidth]{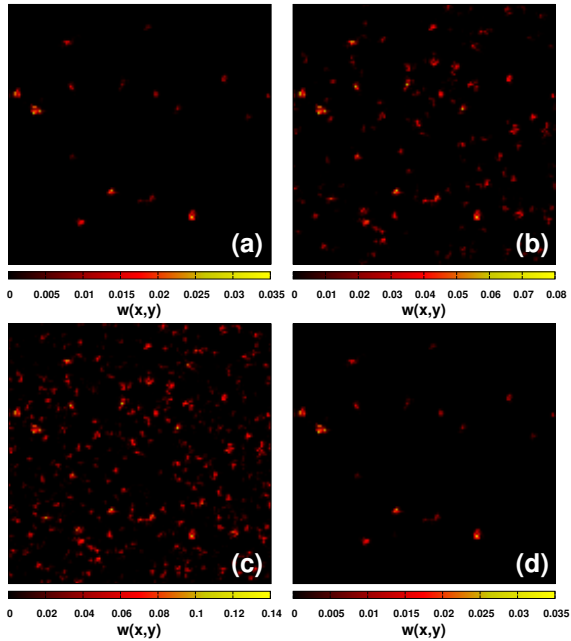}
\end{center}
\vglue -0.3cm
\caption{\label{fig2}Anderson attractor in the DINAP model for a 2D lattice with a linear size $N=128$, a disorder strength $W=8$, $\beta=\sigma=1$, and $\eta=0.1$.
The different panels show the lasing power distribution $w\left(x,y\right)=\left|A_{x,y}\right|$ for the pumping strength
$\alpha=0.09$ (a,d),
$\alpha=0.11$ (b),
and
$\alpha=0.13$ (c).
The lasing power distributions are shown after an evolution time
$t_e=10^4$ (a,b,c)
and
$t_e=10^5$ (d).
The panel (d) shows the attractor stability for long evolution times. The color bars give the values of $w(x,y)$.
The $w(x,y)$ distributions are averaged over a time interval $\delta t = 10^3$.
}
\end{figure}

Typical distributions of the lasing power 
$w(x,y) = |A_{x,y}|^2$ 
on the 2D lattice are shown in Fig.~\ref{fig2}
for different values of the activation strength $\alpha$.
The results show a significant increase of the number of 
lasing attractors with the growth of the $\alpha$-parameter
(see Fig.~\ref{fig2} a,b,c panels).
The Anderson attractor is the steady state of the system since once established, e.g. at $t_e=10^4$ (see panel a), it continue for longer times, e.g. at $t=10^5$ (the lasing power distributions are the same in a and d panels).
The results are shown for 
a typical initial field distribution with random amplitudes
$A_{x,y}$ described in the previous section;
we numerically check that any choice of other
random configurations $A_{x,y}$ does not change the 
average lasing distribution.

\begin{figure}[t]
\begin{center}
\includegraphics[width=0.48\textwidth]{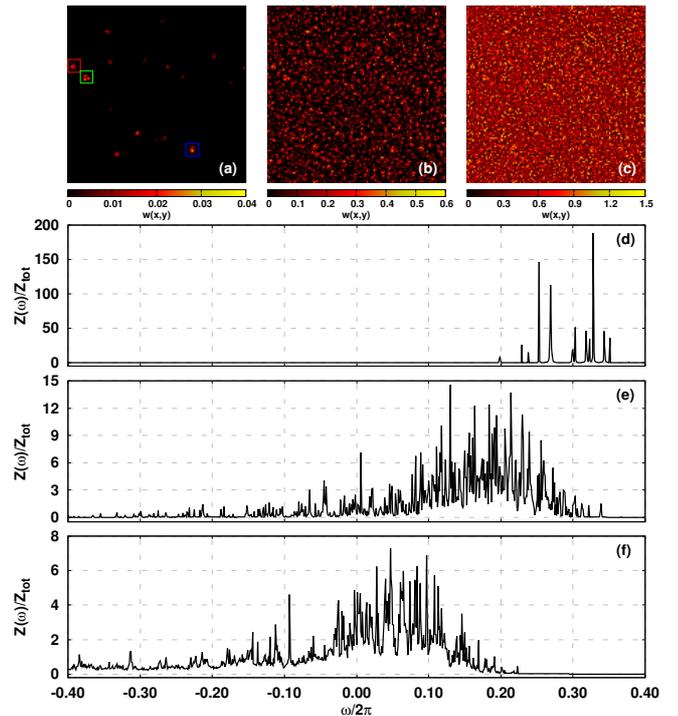}
\end{center}
\vglue -0.4cm
\caption{\label{fig3}Spectrum $Z\left(\omega\right)$ of random lasing in the DINAP model for a 2D lattice with a linear size $N=128$, a disorder strength $W=8$, $\beta=\sigma=1$, and $\eta=0.1$ (same parameters as in Fig.~\ref{fig2}). The pumping strength is 
$\alpha = 0.09$ (a,d),
$0.31$ (b,e),
and
$0.91$ (c,f).
The (a,b,c) panels show the lasing power distributions
$w\left(x,y\right)$
and
the (d,e,f) panels show the spectrum $Z\left(\omega\right)$ of the random lasing.
The integral of the spectral lasing power is
$Z_{tot} \simeq 8.7\times10^{-8}$ (a,d),
$6.0\times10^{-5}$ (b,e),
and
$4.5\times10^{-4}$ (c,f). 
Initial conditions and random realizations are the same as in Fig.~\ref{fig2}(a,d).
}
\end{figure}

Using the fast Fourier transform, we determine the spectrum of the random lasing
defined as $Z(\omega) = \left< \left|\int dt\, A_{x,y}\left(t\right) \exp\left(-i \omega t\right)\right|^2 \right>$ 
where the $\left<\right>$-brackets denote the averaging over the whole lattice space.
We also compute the integral of the spectral power of the random lasing
$Z_{tot}= \int d\omega\, Z\left(\omega\right)$.
The spectrum $Z\left(\omega\right)$ of the random lasing is shown in Fig.~\ref{fig3}
together with the lasing power distribution $w\left(x,y\right)$ over the lattice. 
For small activation strength $\alpha =0.09$, the lasing spectrum is composed of
well separated strong frequency peaks (Fig.~\ref{fig3}d). The space distribution (Fig.~\ref{fig3}a) indicates
that these frequency peaks are generated by well separated lattice cells (clusters).
We call this regime a regime of random lasing clusters. 
With the increase of the pumping strength $\alpha =0.31$, the lasing spectrum becomes 
rather broad even if there are still a couple of dominant
strong frequency peaks emerging from a quasi-continuum spectral component (Fig.~\ref{fig3}e).
Over the lattice, see Fig.~\ref{fig3}b, there are more and more lasing sites as $\alpha$ increases. 
At stronger pumping strength, e.g. $\alpha=0.91$, the lasing spectrum becomes almost continuous (Fig.~\ref{fig3}f), and over the lattice, almost all the sites are lasing (Fig.~\ref{fig3}c).

\begin{figure}[t]
\begin{center}
\includegraphics[width=0.48\textwidth]{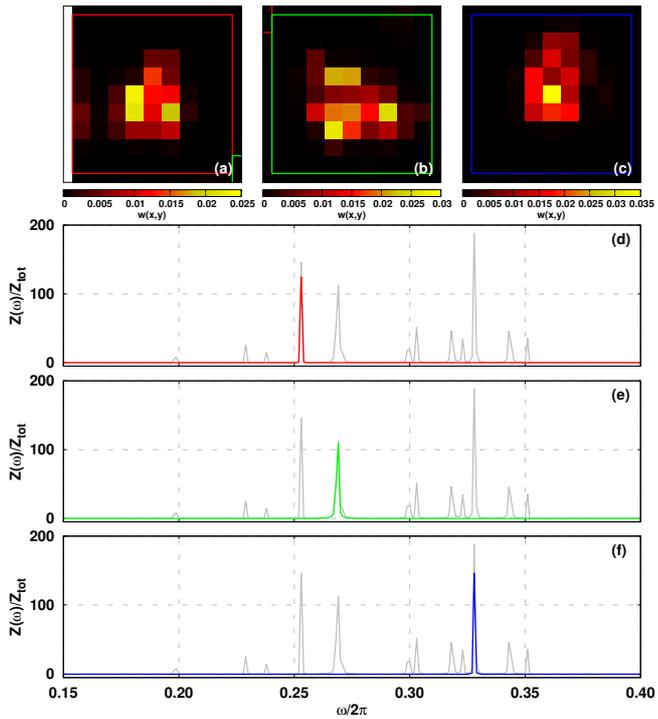}
\end{center}
\vglue -0.4cm
\caption{\label{fig4}The lasing clusters present in the red, green and blue squares in the Fig.~\ref{fig3}a are 
shown in the a, b and c panels, respectively.
The frequencies activated by these clusters 
are shown by the red, green and blue curves in the d, e and f panels, respectively. The spectrum $Z\left(\omega\right)$ of the whole attractor (Fig.~\ref{fig3}d) is drawn in the background of the d, e and f panels. Here, the random lasing spectrum $Z\left(\omega\right)$ is normalized by the integral $Z_{tot}\simeq8.7\times10^{-8}$.
}
\end{figure}

To demonstrate that indeed spectral peaks are generated by 
specific isolated clusters, we select three groups of sites for the small activation strength $\alpha=0.09$ 
delimited by the red, green, and blue color squares in Fig.~\ref{fig3}a.
In Fig.~\ref{fig4}, we superimpose the lasing spectrum of each of the 3 selected clusters onto the lasing spectum obtained for the whole lattice.
Fig.~\ref{fig4} results clearly show
that these 3 selected clusters generate well isolated spectral 
peaks of lasing.

Of course, for another random realization
of the on-site energies $E_{x,y}$,
and for small, moderate and strong activation strengths
$\alpha=0.09, 0.31, 0.91$,
the location of clusters are different
 but the global picture of lasing is
similar to those shown in Figs.~\ref{fig2}~and~\ref{fig3}
which describe a generic situation.

\begin{figure}[t]
\begin{center}
\includegraphics[width=0.48\textwidth]{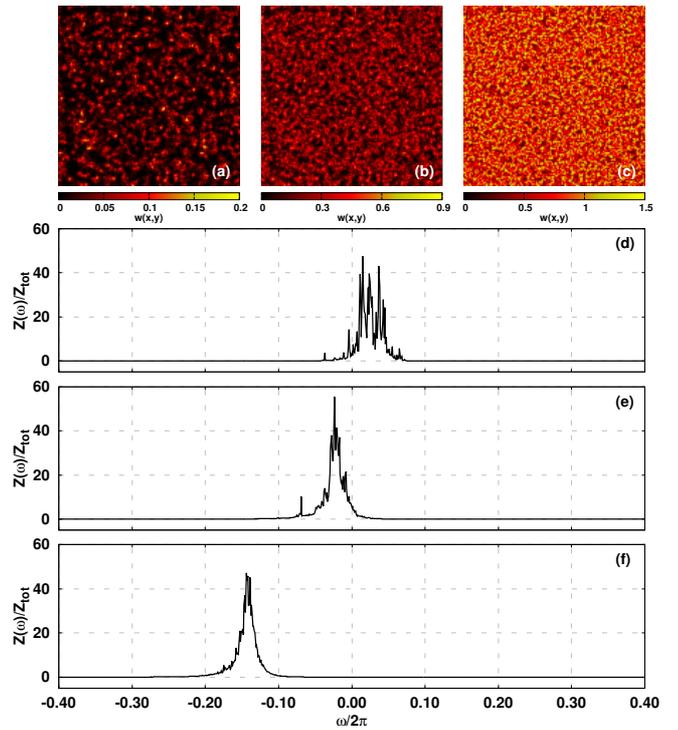}
\end{center}
\vglue -0.4cm
\caption{\label{fig5}
Same as in Fig.~\ref{fig3} but for a disorder strength
$W=3$. Here,
$Z_{tot} \simeq 2.41\times10^{-5}$ for $\alpha=0.09$ (a,d),
$1.9\times10^{-4}$ for $\alpha=0.31$ (b,e),
and
$7.5\times10^{-4}$ for $\alpha=0.91$ (c,f).
}
\end{figure}

In the above presented figures, we considered the case when 
the linear system (ie, the linear modes of the corresponding
Anderson model) has well localized eigenstates
with a localization support being significantly smaller 
than the linear system size $N$ 
(see the typical eigenstate characteristics for $W=8$ at \cite{lagesbls}).
This regime is characterized by narrow peaks of lasing
spectrum generated by isolated localized clusters.
It is also interesting to consider the opposite case
when linear modes have a support being comparable to 
the linear system size thus corresponding to
the metallic regime. For $W=3$, we have approximately such a regime
according to the results presented in \cite{lagesbls}.
The power distribution $w(x,y)$ over the lattice and the
lasing spectrum $Z\left(\omega\right)$ for such a case are shown in Fig.~\ref{fig5}.
In this metallic regime, even for a small activation strength 
$\alpha = 0.09$, we have a broad spatial power distribution of lasing;
the lasing spectrum is quasi-continuous.
For strong activation strength $\alpha=0.91$,
almost all the lattice sites are lasing. The lasing spectrum 
has a structure being similar to the one for
small $\alpha = 0.09$ but with a larger number of spectral peaks.
%??COMPARE LASING AT W=8 (fig3) and W=3 fig5;
%use the same Z_{tot} normalization as the global definision
%and NOT a different one as now in fig5
%I assume that lasing peaks are stronger
%at localized regime => check??
The important feature of the metallic regime at $W=3$
is that all spectral peaks visible in the localized regime at $W=8$
are replaced by a quasi-continuous broad distribution.
Thus, the localized regime is better adapted to
a narrow spectral line of lasing.

The integrated lasing power,
taking the same pumping and dissipation parameters,
is globally higher for the metallic phase regime
(see e.g. Fig.~\ref{fig5}cf with $W=3$, $Z_{tot}\simeq 7.5\times10^{-4}$)
than for the localized phase regime
(see e.g. Fig.~\ref{fig3}cf with $W=8$, $Z_{tot} = 4.5\times10^{-4}$).
We attribute this to the fact
that more sites contribute to
the lasing in the metallic phase
due to delocalized eingestates of the linear
Anderson model.

\section{Discussion} 
\label{sec4}

We introduced a mathematical 2D DINAP model
to describe specific features of the
random lasers: lasing above a certain threshold,
pronounced spectral  lasing peaks,
lasing clusters. Our numerical
analysis shows that
these features are well described
by the DINAP model.
The important element of the model
is that in the linear regime
without nonlinearity and dissipation
it is reduced to the 2D Anderson model
with localized modes at strong disorder
and delocalized ones at weak disorder
when the finite system size
becomes comparable with 2D localization length
(in the infinite system). In the localized phase,
above the critical pumping strength $\alpha$,
the spectrum of lasing
is composed of narrow spectral lines.
These lines emanate from localized isolated clusters
located in a media where a nonlinear dissipative
dynamics leads to isolated Anderson attractors.
With the increase of the pumping strength, the lasing
peaks are still present but a global
envelope appears corresponding to lasing from
a large number of connected or disconnected
clusters. Globally, an increase
of the pumping strength leads to a broader
lasing spectrum. We find that such an effect
is rather natural since with the increase of the pumping,
the nonlinear frequency corrections become higher.
In the metallic regime, the peaks are significantly
less visible even if only slightly above the threshold pumping
and at higher strength of pumping,
the lasing spectrum gets a form of
a smooth envelope. We attribute this feature to a delocalized structure of
linear modes where nonlinear frequency corrections
at high pumping get contribution from
many lattice sites on which are located the delocalized linear modes.

In the localized phase, our results show that a percolation transition takes place
from a localized lasing clusters regime to a delocalized regime
of lasing from many lattice sites when the strength of the pumping
significantly increases above the threshold value.
There is a number of interesting questions about such a percolation:
What is the critical percolation threshold?
How it depends on the system parameters?
Is there a global synchronization of lasing, like the Kuramoto transition \cite{sync}?
Is there a superradiance in such a synchronized phase?
We think that the answers to these questions
can be obtained in the further investigations
of the DINAP model.

% to discuss 
%
%- percolation
%
%- global synchronization
%
%- strong narrow peaks in metallic regime ?

\begin{acknowledgments}
This research has been partially supported through the grant
NANOX $N^\circ$ ANR-17-EURE-0009 (project MTDINA) in the frame 
of the {\it Programme des Investissements d'Avenir, France}.
\end{acknowledgments}

%%%%%%%%%%%%%%%%%%%%%%%%%%%%%%%%%%%%%%%%%%%%%%%%%%%%%%%%%

\end{document}